\documentclass[twocolumn,superscriptaddress,amsmath,amssymb]{revtex4-2}
\usepackage{color}
\usepackage{graphicx}% Include figure
\usepackage{subfigure}% subplot
\usepackage{booktabs}
\usepackage{dcolumn}% Align table columns on decimal point
\usepackage{bm}% bold math
\usepackage{amsmath}%more math font
\usepackage{verbatim}%multi-line "comment" environment
\usepackage{lineno}
\newcommand{\beq}{\begin{eqnarray}}
\newcommand{\eeq}{\end{eqnarray}}

\usepackage[normalem]{ulem}

\begin{document}

\title{Gardner transition coincides with the emergence of jamming scalings in hard spheres and disks}

\author{Qi Wang}
\affiliation{Science and Technology on Surface Physics and Chemistry Laboratory, Mianyang, Sichuan 621908, China}

\author{Deng Pan}
\email{dengpan@mail.itp.ac.cn}
\affiliation{Institute of Theoretical Physics, Chinese Academy of Sciences, Beijing 100190, China}

\author{Yuliang Jin}
\email{yuliangjin@mail.itp.ac.cn}
\affiliation{Institute of Theoretical Physics, Chinese Academy of Sciences, Beijing 100190, China}
\affiliation{School of Physical Sciences, University of Chinese Academy of Sciences, Beijing 100049, China}
\affiliation{Center for Theoretical Interdisciplinary Sciences, Wenzhou Institute, University of Chinese Academy of Sciences, Wenzhou, Zhejiang 325001, China}

\date{\today}

\begin{abstract}
The Gardner transition in structural glasses is characterized by full-replica symmetry breaking of the free-energy landscape and the onset of anomalous aging dynamics due to marginal stability. Here we show that this transition also has a structural signature in finite-dimensional glasses consisting of hard spheres and disks. By analyzing the distribution of inter-particle gaps in the simulated static configurations at different pressures, we find that  the Gardner transition coincides with the emergence of two well-known jamming scalings in the gap distribution, which enables the extraction of a structural order parameter.
The jamming scalings reflect a compressible effective force network formed by contact and quasi-contact gaps, 
 while non-contact gaps that do not participate in the effective force network are incompressible. Our results suggest that the Gardner transition in hard-particle glasses is a precursor of the jamming transition. The proposed structural signature and order parameter provide a convenient approach to detecting the Gardner transition in future granular experiments.      
\end{abstract}

\maketitle

{\bf Introduction.} 
The  spin glass transition in mean-field models, such as the Sherrington–Kirkpatrick model~\cite{sherrington1975solvable}, is characterized by  full-replica symmetry breaking (full-RSB), a solution initially provided by Parisi~\cite{parisi1979infinite, parisi1980sequence, mezard1987spin, parisi2023nobel}, and later proved rigorously~\cite{guerra2001sum, guerra2003broken, talagrand2006parisi, panchenko2013parisi}. The counterpart of the full-RSB spin glass transition in structural glasses is called the {\it Gardner transition}~\cite{gardner1985spin,
berthier2019gardner, urbani2023gardner}, predicted recently by  mean-field replica theory (MFRT) calculations~\cite{charbonneau2014exact, charbonneau2014fractal, parisi2020theory}. The evidence of the density(pressure)-controlled Gardner transition (crossover) in finite-dimensional structural glasses has been consistently observed in simulations of hard particles~\cite{berthier2016growing, jin2017exploring, jin2018stability, liao2019hierarchical, charbonneau2019glassy, liao2023dynamic, li2021determining}, and in experiments of granular matter~\cite{seguin2016experimental, kool2022gardner, xiao2022probing}. However, numerical and experimental searching for the temperature-controlled Gardner transition has  mostly failed in soft-potential systems~\cite{scalliet2017absence, seoane2018low,  albert2021searching}, despite that the MFRT  universally identifies the Gardner transition in corresponding models~\cite{scalliet2019marginally}. 

In this study, we aim to answer two questions. (i) Although according to the MFRT, the mean-field spin glass transition and the Gardner transition belong to the same full-RSB universality class, they can nevertheless be different in some aspects. In principle, (non-crystalline) structural ordering of particle arrangement can emerge in structural glasses, which by definition is irrelevant in spin glasses. Is the Gardner transition accompanied with structural indicators? (ii) Why is it more robust to detect the density-controlled Gardner transition in hard particles than the temperature-controlled Gardner transition in soft particles? 
As shown by this study, the answers to both questions have a common origin. Hard particles encounter a jamming transition upon compression, at which the static distribution of inter-particle gaps  is characterized by universal power-law scalings. The onset of the jamming scalings coincides with the Gardner transition in hard particles, providing a structural interpretation of the Gardner transition. On the other hand, jamming scalings do not appear in  soft particles under constant-density cooling, since they can not jam asymptotically. In other words, our results suggest that the finite-dimensional Gardner transition is a ``precursor'' of the jamming transition in hard particles.

\begin{figure}[!htbp]
  \centering
  \includegraphics[width=0.7\linewidth]{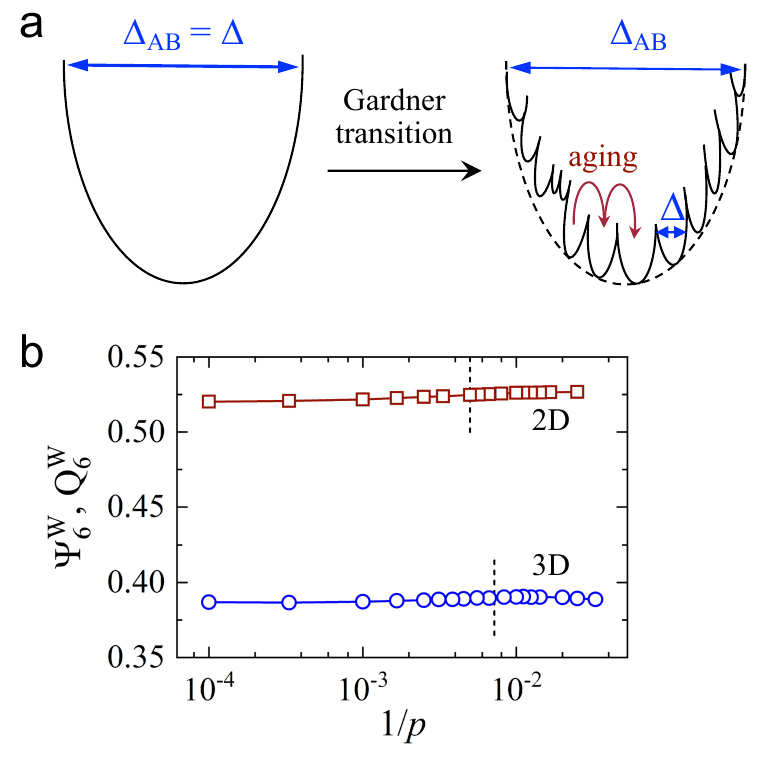}
  \caption{{\bf Thermodynamic and dynamic interpretations of the Gardner transition.} (a) Schematic free-energy landscapes below and above the Gardner transition. The split of the meta-stable basin is characterized by the 
  caging order parameter $\Delta_{AB} - \Delta$, and aging dynamics. (b) Weighted  bond orientational (structural) order parameters, $\Psi_{6}^{\rm w}$ in 2D and $Q_{6}^{\rm w}$ in 3D, do not capture the Gardner transition. Vertical dashed segments indicate $p_{\rm G}\approx 200$ in 2D and  $p_{\rm G}\approx 139$ in 3D, respectively.
  } 
\label{fig:schematic}
\end{figure}

The existing understanding of  the Gardner transition has focused on thermodynamic and dynamic aspects (see Fig.~\ref{fig:schematic}a).
(i) From the thermodynamic viewpoint, at the  Gardner transition a meta-stable glass basin splits into many hierarchical sub-basins. The typical sizes of the meta-stable basin and the minimum sub-basin are characterized by caging order parameters $\Delta_{\rm AB}$ and $\Delta$. Below the Gardner transition pressure $p_{\rm G}$, $\Delta_{\rm AB} = \Delta$, and above $p_{\rm G}$, $\Delta_{\rm AB} > \Delta$. Thus the difference $\Delta_{\rm AB} - \Delta$ serves as an order parameter (called {\it caging order parameter}) of the Gardner transition, which has been used to locate $p_{\rm G}$ in previous simulations of hard spheres~\cite{charbonneau2015numerical, berthier2016growing} and hard disks~\cite{liao2019hierarchical, liao2023dynamic}, and in experiments of granular glasses~\cite{seguin2016experimental, kool2022gardner, xiao2022probing}. 
(ii) From the dynamic viewpoint, the change of the  phase space structure at $p_{\rm G}$ is accompanied with the onset of aging dynamics. For $p<p_{\rm G}$, the system has reached 
{\it restricted equilibrium} within the meta-stable basin
~\cite{berthier2016growing}. However, the system becomes out of the restricted equilibrium for $p>p_{\rm G}$. In other words, the Gardner transition  breaks the ``ergodicity'' in a glass state. 
Here the restricted equilibrium means that the mean-squared displacement (MSD) has reached a stationary plateau that is independent of the waiting time, and meanwhile the $\alpha$-relaxation occurs at a much later time that is inaccessible in simulations. Note that such a clear separation of $\beta$- and $\alpha$-time scales is possible in ultra-stable glasses~\cite{berthier2016growing}.

It was unknown whether the Gardner transition can be captured by  any characteristic change in the static structure of the glass configuration. Conventional structural measures, such as the weighted bond orientational order parameters~\cite{mickel2013shortcomings}, fail to capture this transition (see Fig.~\ref{fig:schematic}b and Appendix A). 
In this study we focus on the distribution $\rho(x)$ of inter-particle gaps  $x$ computed for each pair of adjacent particles in the Voronoi tessellation (see Appendix B).  
Our motivation  is as follows. Near the jamming limit ($p_{\rm J} = \infty$), the power-law scalings of the small-$x$ distribution have been established~\cite{charbonneau2012universal, charbonneau2014fractal, charbonneau2015jamming}, where $\rho(x) \sim g(r-1)$ with $g(r)$ the pair correlation function. In equilibrium states ($p=p_{\rm eq}$),  the gap distribution $\rho(x)$ should follow a generalized gamma distribution, suggested by empirical observations~\cite{sastry1998free, hoover1979exact}.  Thus the behavior of $\rho(x)$ at the equilibrium and jamming limits 
can be well described by existing knowledge. Our task is to understand how the two are connected in the intermediate regime, where the Gardner transition occurs ($p_{\rm eq} < p_{\rm G} < p_{\rm J}$).

We employ the swap Monte-Carlo algorithm~\cite{berthier2016equilibrium} to prepare parent liquid configurations of $N$ polydisperse particles interacting via hard-core potentials in two (2D) and three (3D) dimensions (see Appendix C), equilibrated at a (reduced) pressure $p_{\rm eq}$ above the  mode-coupling theory (MCT) crossover pressure $p_{\rm MCT}$. Here, $p_{\rm eq} = 31, p_{\rm MCT} \approx 25$ in 2D, and $p_{\rm eq} = 30, p_{\rm MCT} \approx 23$ in 3D.
Ultra-stable hard-sphere glasses are generated by rapidly compressing the system from $p_{\rm eq}$ to a target pressure $p$, using the Lubachevsky–Stillinger molecular dynamics (MD) algorithm (without swapping)~\cite{lubachevsky1990geometric} (see Appendix D). The $\alpha$-relaxation time $\tau_\alpha$ of these ultra-stable states is significantly beyond the maximum MD simulation time, and thus large-scale  structural rearrangements  are  suppressed during the compression. 
For the above setup, previous studies have estimated $p_{\rm G}\approx 200$ in 2D~\cite{liao2019hierarchical} 
and $p_{\rm G}\approx 139$ in 3D~\cite{berthier2016growing}   based on the caging order parameters and the onset of aging dynamics. A supervised machine learning approach gives a consistent value of $p_{\rm G}\approx 140$ in 3D~\cite{li2021determining}.

\begin{figure}[!htbp]
  \centering
  \includegraphics[width=\linewidth]{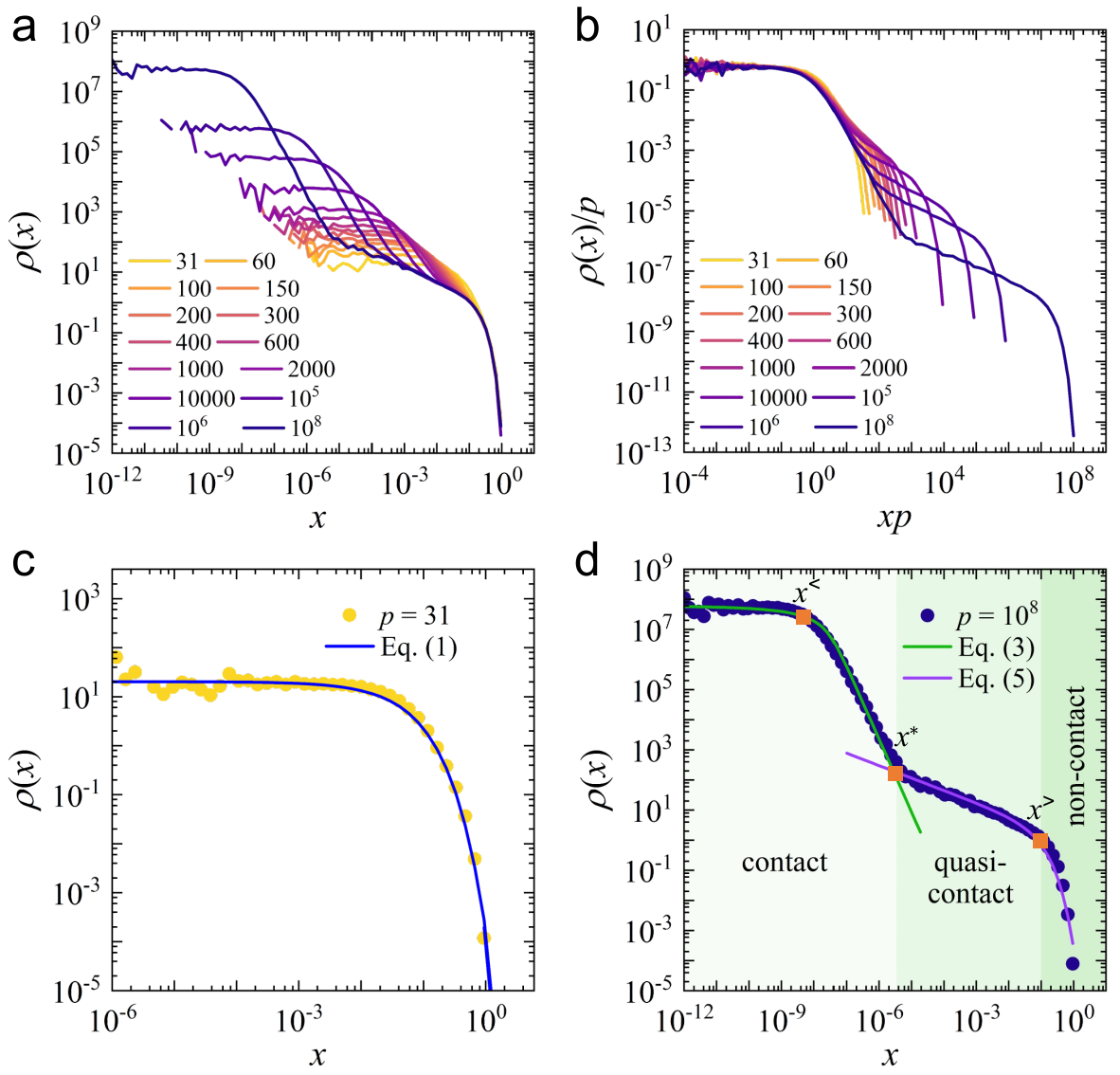}
  \caption{{\bf Gap distributions in 2D.} (a) Original and (b) $p$-rescaled distribution $\rho(x)$ at different pressures.
  (c) Equilibrium $\rho(x)$ fitted to Eq.~(\ref{eq:Peq}). (d) Near-jamming $\rho(x)$ fitted to Eqs.~(\ref{eq:PJc}) and~(\ref{eq:Pstar}).
  } 
\label{fig:PDF}
\end{figure}

{\bf Gap distributions  near equilibrium and jamming limits.} Figure~\ref{fig:PDF}a shows  $\rho(x)$  at different $p$ in 2D, ranging from $p_{\rm eq}=31$ to  $p=10^8$ near  $p_{\rm J} = \infty$. 
The equilibrium gap distribution  $\rho_{\rm eq}(x)$ at $p = p_{\rm eq}$ can be described empirically by a generalized gamma distribution, 
\beq
\rho_{\rm eq}(x) = k \, x^{\beta} \exp\left(- a \, x^\gamma \right).
\label{eq:Peq}
\eeq
where $a$, $\beta$ and $\gamma$ are fitting parameters, and $k$ is a normalization constant (see Fig.~\ref{fig:PDF}c). The same form has been proposed to describe the distribution of free volumes, $\rho_{\rm eq}(v_{\rm f}) \sim v_{\rm f}^{\beta} \exp\left(- a \, v_{\rm f}^\gamma \right)$, in equilibrium hard-sphere fluids, with $\beta = 0.28-0.35$  and $\gamma = 0.55-0.45$ for the range of densities investigated~\cite{sastry1998free}. The $\rho_{\rm eq}(x)$ of  our system can be well fitted by Eq.~(\ref{eq:Peq}) by setting $\beta = 0$, which gives $\gamma = 0.80$ from fitting.

{Near the jamming limit $p \to \infty$, $\rho(x)$ consists of three regimes (see Fig.~\ref{fig:PDF}d)~\cite{charbonneau2012universal}: a {\it contact} regime denoted by $\rho_{\rm c}(x) \equiv \rho(x<x^*)$, a  {\it quasi-contact} regime by $\rho_{\rm qc}(x) \equiv \rho(x^*< x < x^>)$, and  a {\it non-contact} regime by $\rho_{\rm nc}(x) \equiv \rho(x > x^>)$.
The characteristic gaps $x^<$, $x^*$ and $x^>$ correspond to the three kink points as indicated in Fig.~\ref{fig:PDF}d.}

(i) In the contact regime,  the distributions $\rho_{\rm c}(x, p)$ at different $p$  obey the scaling form  $\rho_{\rm c}(xp)$ (Fig.~\ref{fig:PDF}b), implying that these gaps $x \sim 1/p$ would disappear as $p \to \infty$. In other words, the pair of particles forming the gap would be in contact at jamming. 
The scaling $x \sim 1/p$ follows  the hard-sphere equation of state near jamming,  derived from the free volume theory~\cite{salsburg1962equation}.
For $x > x^<$, $\rho_{\rm c}(x)$ decays as a power-law,
\beq
\rho_{\rm c} (x) \sim x^{-2-\theta},
\label{eq:scaling_Pc}
\eeq
which is related to the jamming scaling of the weak force distribution $\rho(f) \sim f^\theta$~\cite{charbonneau2012universal, lerner2013low} through a Laplace transform, with $\theta = 0.42311$ predicted by the mean-field replica theory~\cite{charbonneau2014fractal, charbonneau2014exact} and confirmed by finite-dimensional simulations~\cite{charbonneau2015jamming}.  
In the limit $x \to 0$, $\rho_{\rm c}(x)$ converges to a constant plateau. 
{To describe  $\rho_{\rm c}(x)$ in this regime, we propose the following form,
\beq
\rho_{\rm c}(x) = b_1\left[ 1 - \frac{1}{\left(1+b_2\,x^{-2-\theta} \right)^{b_3}} \right],
\label{eq:PJc}
\eeq
where $\theta = 0.42311$ is fixed, while $b_1, b_2$ and $b_3$ are fitting parameters (see Fig.~\ref{fig:PDF}d).} The asymptotic behavior of Eq.~(\ref{eq:PJc}) at large $x$ recovers the well-known jamming scaling Eq.~(\ref{eq:scaling_Pc}).

{
(ii) In the quasi-contact regime,  the free-volume scaling $x \sim 1/p$ is not satisfied (the data in this regime do not collapse in Fig.~\ref{fig:PDF}b). 
The contact and quasi-contact regimes are separated by the characteristic gap size $x^*$.
In the limit $p \to \infty$, the contact regime $\rho_{\rm c}(x)$ disappears, since $x^* \to 0$ (see Fig.~\ref{fig:PDF}a).  It means that at jamming, all contact gaps are closed, while quasi-contact particles remain separated. For $x^*< x < x^>$, $\rho_{\rm qc}(x)$ displays another jamming scaling~\cite{charbonneau2015jamming, jin2021jamming},
\beq
\rho_{\rm qc}(x)  \sim x^{-\alpha},
\label{eq:scaling_Pqc}
\eeq
with $\alpha = 0.41269$ given by the  mean-field theory~\cite{charbonneau2014fractal, charbonneau2014exact}.}

(iii) In the non-contact regime, the distribution is independent of $p$. It can be shown that (see below) the particles forming these gaps do not collide during constant-volume dynamics. {The quasi-contact and non-contact gap distributions ($x>x^*$) can be well described by
\beq
\rho_{*}(x) = c_1 \left[ x^{-\alpha} \exp\left(- c_2 \, x^\gamma \right) \right].
\label{eq:Pstar}
\eeq
where $\alpha = 0.41269, \gamma=0.80$ are fixed, and $c_1, c_2$ are fitting parameters (see Fig.~\ref{fig:PDF}d).
With this setting, the small-$x$ behavior of Eq.~(\ref{eq:Pstar}) is consistent with the jamming scaling Eq.~(\ref{eq:scaling_Pqc}), and the large-$x$ part of Eq.~(\ref{eq:Pstar}) agrees with that of the equilibrium behavior Eq.~(\ref{eq:Peq}).}

\begin{figure*}[!htbp]
  \centering
  \includegraphics[width=0.9\linewidth]{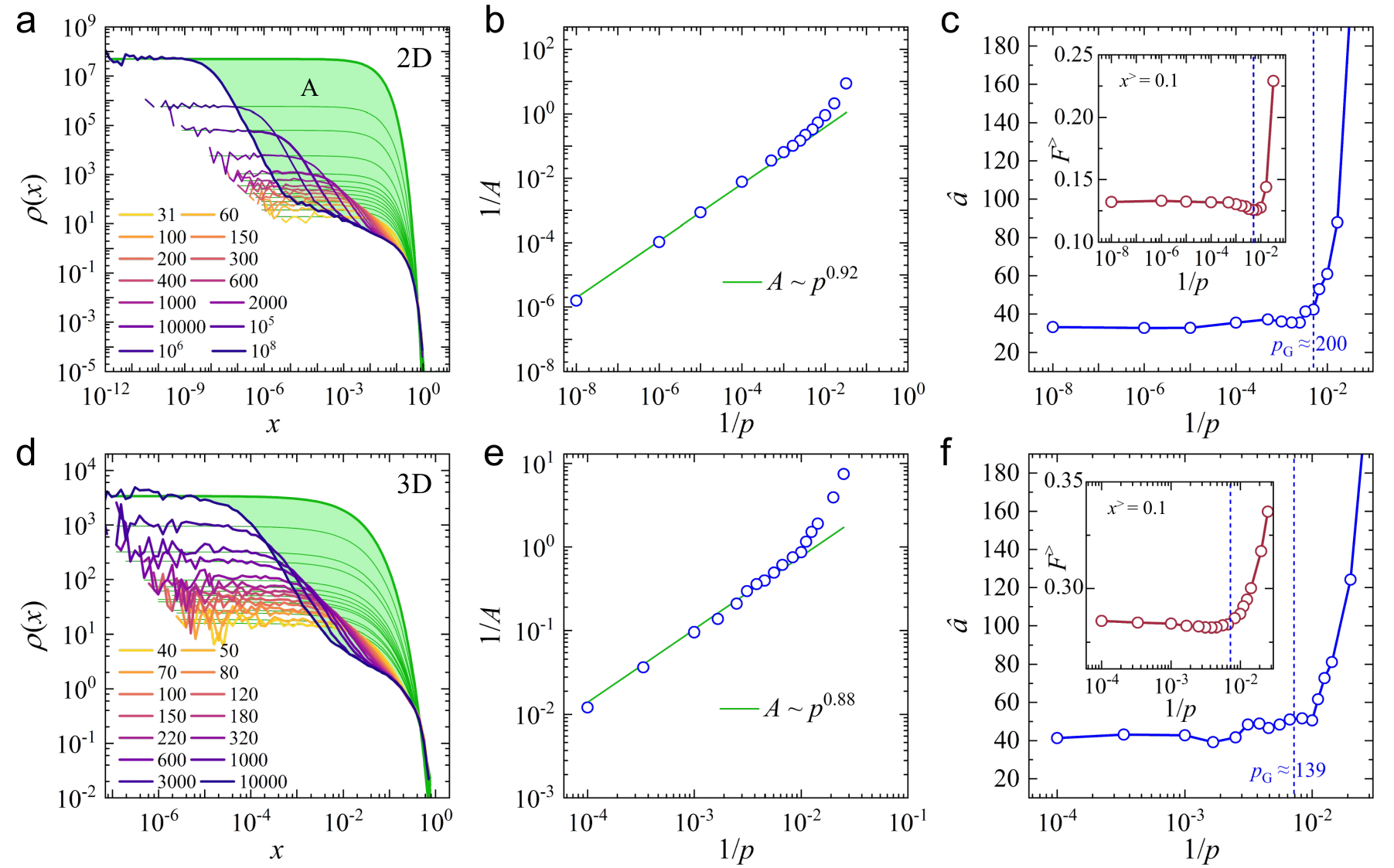}
  \caption{{\bf Structural order parameter of the Gardner transition in (a-c) 2D and (d-f) 3D.}
  In (a,d),  green lines represent the fitting to Eq.~(\ref{eq:Peq}), and the green areas represent $A$ (see Eq.~(\ref{eq:A})).
  (b,e) The large-$p$ data are fitting to Eq.~(\ref{eq:scalingA}).
  (c,f) $\hat{a}$ and (inset) $F^>$ as  functions of $1/p$.
  } 
\label{fig:order_parameter}
\end{figure*}

\begin{figure}[!htbp]
  \centering
  \includegraphics[width=0.9\linewidth]{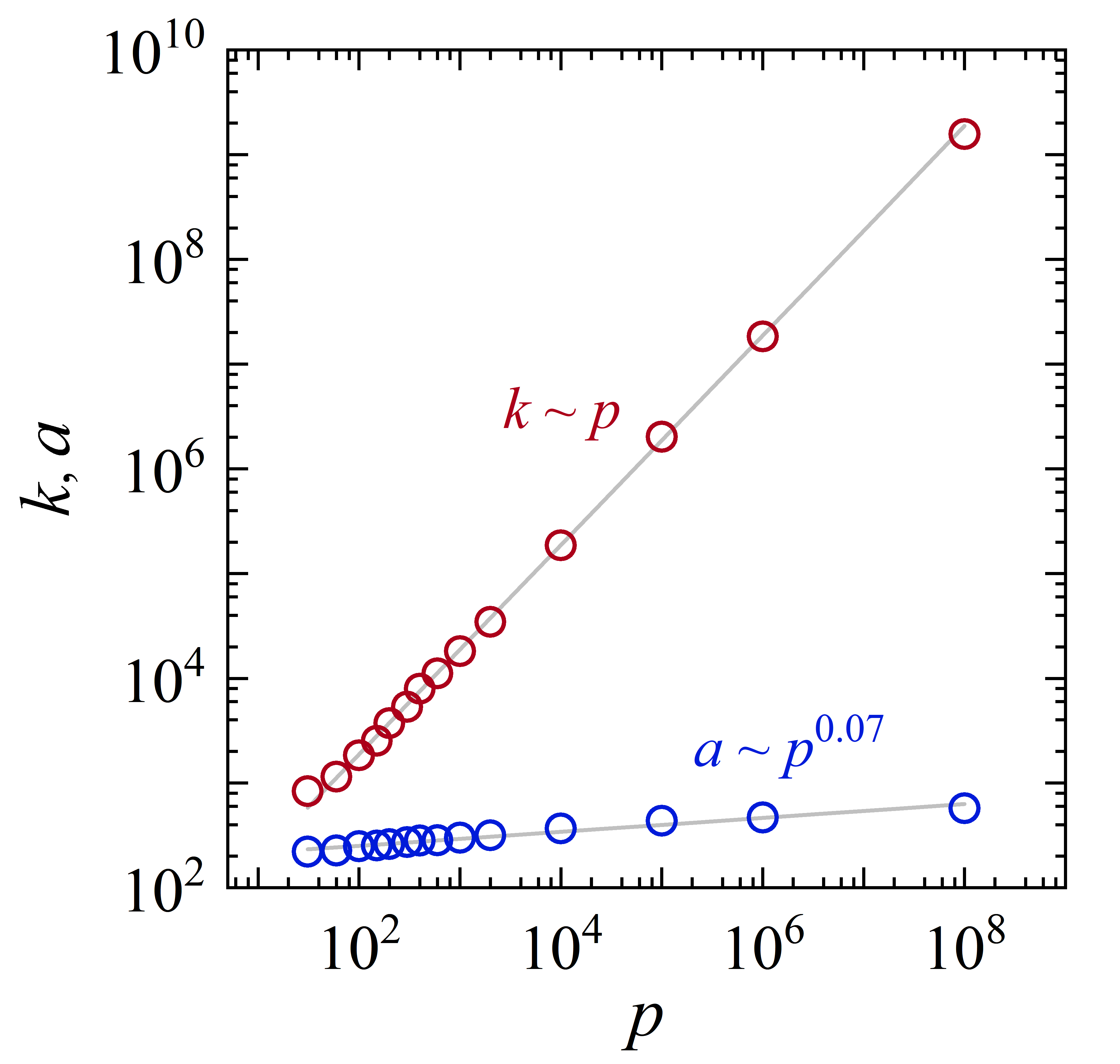}
  \caption{Fitting parameters $k$ and $a$ as  functions of pressure. The solid lines are scaling relations with exponent $1.0$ and $0.07$.} 
\label{fig:rhoeq_parameters}
\end{figure}

{\bf A structural order parameter.}
In the above analyses, three characteristic gap sizes, $x^<, x^*, x^>$, are defined near jamming (Fig.~\ref{fig:PDF}d). Interestingly, if the power-law scaling regime $x \in [x^<, x^>]$ is skipped, the remaining data of $\rho(x)$ can be well fitted by the equilibrium form Eq.~(\ref{eq:Peq}) (with $\gamma=0.80$ fixed, $k$ and $a$ treated as  fitting parameters), as if the small gaps with $x < x^<$ and the large gaps with $x > x^>$  were still in an equilibrium state (see Fig.~\ref{fig:PDF}(a,d)). This observation inspires us to compute the area $A$ (the green area in Fig.~\ref{fig:PDF}(a,d)) between the ``fake'' equilibrium distribution $\rho'_{\rm eq}(x)$ obtained from fitting and the actual $\rho(x)$,
\beq
A = \int_{0}^\infty \left[ \rho'_{\rm eq}(x) -  \rho(x) \right ]dx = \int_{0}^\infty \rho'_{\rm eq}(x) dx -1.
\label{eq:A}
\eeq
 In the equilibrium limit $p \to p_{\rm eq}$, $A 
\to 0$ by construction. At larger $p$, Eqs.~(\ref{eq:Peq}) and~(\ref{eq:A}) give 
$A = \int_{0}^\infty ke^{-a x^\gamma}dx -1 = \frac{k}{\gamma}  a^{-1/\gamma} \Gamma\left( 1/\gamma \right) -1$,
where  $\Gamma(x)$ denotes the gamma function. Thus the  scaling of $A$  is determined by $A \sim ka^{-1/\gamma}$. Our data suggest that, $k \sim p$ and $a \sim p^{\nu}$ with $\nu \approx 0.07$ 
 (see Fig.~\ref{fig:rhoeq_parameters}), which give, 
\beq
A \sim p^\kappa,
\label{eq:scalingA}
\eeq
where $\kappa = 1-\nu/\gamma \approx 0.9$ (see Fig.~\ref{fig:order_parameter}(b,e)).

We propose the scaled inverse area, $\hat{a} = p^\kappa/A$, as a structural order parameter of the Gardner transition. As discussed above, $\hat{a}$ is a constant at large $p$ and diverges at $p_{\rm eq}$. Remarkably, the deviation point from the large-$p$ plateau coincides nicely with  $p_{\rm G}$  estimated independently from the caging order parameters in previous studies in both 2D and 3D~\cite{berthier2016growing, liao2019hierarchical} (see Fig.~\ref{fig:order_parameter}(c,f)). 
It should be emphasized that, unlike the caging order parameters that are obtained from vibrational dynamics, $\hat{a}$ is defined purely based on  static information extracted from instantaneous configurations. We also find that $\hat{a}(p)$ is independent of the system size $N$ and the waiting time $t_{\rm w}$ (the system is relaxed for a time $t_{\rm w}$ after quenched to the target $p$), as shown in Appendix F.

To understand why $\hat{a}$ can capture the Gardner transition, we should  clarify the behavior of large and small gaps near jamming. 
The $p$-dependence of  $x^*$, $x^<$ and $x^>$ can be derived by scaling analyses. From the $\rho(x)$ data in Figs.~\ref{fig:PDF} and~\ref{fig:order_parameter}, we find that  $x^< \sim p^{-1}$ and $x^> \sim {\rm constant}$. Matching the two power-law scalings (Eqs.~\ref{eq:scaling_Pc} and~\ref{eq:scaling_Pqc}), $\rho(x^*) \sim (x^*)^{-2-\theta}$ and $\rho(p x^*)/p \sim (p x^*)^{-\alpha}$ at $x^*$, gives $x^* \sim p^{-\frac{1+\theta}{2 + \theta -\alpha}}$, where $\frac{1+\theta}{2 + \theta -\alpha} \approx 0.71 $ using $\alpha = 0.41269$ and $\theta = 0.42311$. {These scalings are confirmed by simulation data in Appendix G.} 
Interestingly, the fraction of large gaps, $F^> = \int_{x^>}^\infty \rho(x) dx$, behaves similarly to $\hat{a}(p)$ (see insets of  Fig.~\ref{fig:order_parameter}(c,f)). It suggests that the number of large gaps is reduced upon compression before the Gardner transition ($p<p_{\rm G})$, and  becomes nearly conserved for $p>p_{\rm G}$.

Based on the above discussion, we derive a structural interpretation of the Gardner transition. Near equilibrium, $\rho(x)$ follows a single scale distribution Eq.~(\ref{eq:Peq}), where small and large gaps are reduced simultaneously by compression, similar to the situation in a homogeneous affine deformation.  Near jamming, the small gaps have to vanish following the free-volume theory ($x^< \sim p^{-1}$), while the large gaps are ``locked'' ($F^>\sim x^> \sim {\rm constant}$)
{because they are incompressible}
(see discussions below) --  consequently two jamming power-law scalings Eqs.~(\ref{eq:scaling_Pc}) and~(\ref{eq:scaling_Pqc}) appear for intermediate gaps, $x \in [x^<, x^>]$. The incompatibility between the behavior of small and large gaps emerges at $p_{\rm G}$, which corresponds to the Gardner transition.

\begin{figure}[!htbp]
  \centering
  \includegraphics[width=\linewidth]{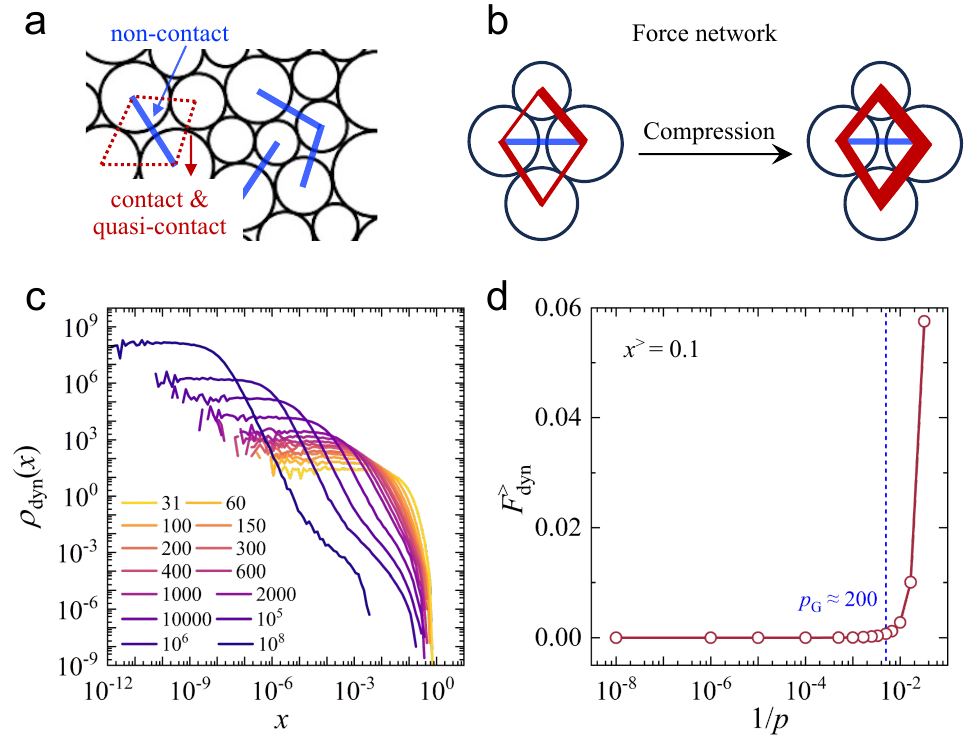}
  \caption{{\bf Effective force network and dynamic gaps.}
  (a) Illustration of non-contact gaps (blue), and contact/quasi-contact gaps (red). (b) The contact and quasi-contact gaps, which form an effective force network (red), are compressible. In this cartoon, the line width is proportional to effective force magnitude. In contrast, the non-contact gaps (blue) are incompressible. (c) Probability distribution $\rho_{\rm dyn}(x)$ of dynamic gaps in 2D. (d) Fraction of large dynamic gaps $F^{>}_{\rm dyn}$.
  }
\label{fig:spatial_distribution}
\end{figure}

{\bf Incompressibility of non-contact gaps in the Gardner phase.}
The remaining question is to understand why  the population $F^>$ of non-contact gaps $x>x^>$ is preserved under compression in the Gardner phase, violating the free-volume scaling $x \sim p^{-1}$. 
The answer is obtained by analyzing the vibrational dynamics. In the Gardner phase $p>p_{\rm G}$, the contact and quasi-contact gaps form an effective force network. However,  the non-contact gaps do not participate in this network, because the neighboring particles forming such gaps do not collide during the time interval the effective force network is stable.  The contact and quasi-contact gaps (or the effective springs) can be compressed with an increasing $p$, but not the non-contact gaps (see Fig.~\ref{fig:spatial_distribution}(a,b)).

Near jamming, a pair of neighboring hard particles interact via an effective force $f \sim x^{-1}$ due to collisions~\cite{brito2006rigidity}. If two neighboring particles do not collide, they do not participate in the effective force network. To examine the dynamic role of gaps, the following simulations and analyses are performed. We first compress the system  to the target $p$ with a particle inflating rate $\Gamma = 10^{-3}$, and then perform constant-volume ($\Gamma = 0$) event-driven MD simulations. After the compression, the system is relaxed for $t_{\rm w} = 10^6$ collisions, and then the next $10^5 \times N$ collisions are analyzed. During the $10^5 \times N$ collisions, no structural relaxation is observed (the MSD is a constant), and  the effective force network is stable.
If two neighboring particles collide during this time interval, then their gap $x$ (in the configuration at $t_{\rm w}$)
is called a {\it dynamic gap}. The distributions $\rho_{\rm dyn}(x)$ of dynamic gaps at different $p$ (in 2D) are presented in Fig.~\ref{fig:spatial_distribution}c. Compared to the static gap distribution Fig.~\ref{fig:order_parameter}a, the difference mainly appears in the large-gap regime.  The percentage of large dynamic gaps, $F^>_{\rm dyn} = \int_{x^>}^\infty \rho_{\rm dyn}(x) dx$,  is shown in Fig.~\ref{fig:spatial_distribution}d. Above $p_{\rm G}$, $F^>_{\rm dyn} = 0$, which means that particles forming these large gaps do not collide at all. 
These gaps do not participate in the effective force network, and thus cannot be compressed or eliminated. This is the reason why the static fraction $F^>$ is a non-zero constant below $p_{\rm G}$ in Fig.~\ref{fig:order_parameter}c-inset. For the same reason, these large gaps are called non-contact gaps. Below $p_{\rm G}$,  $F^>_{\rm dyn}$ rapidly grows, and the non-contact gaps can not be well distinguished from other gaps anymore.

{\bf Discussion.}
This study establishes the structural signature of the Gardner transition in hard spheres and disks. 
For $p<p_{\rm G}$, the effective force network is fragile under compression. Once the system enters the Gardner phase ($p>p_{\rm G}$), it is supported by a stable effective force network. This force network can be compressed, following the scaling $x \sim f^{-1} \sim p^{-1}$,  and the distributions of the contact and quasi-contact gaps in the network display the jamming scalings Eqs.~(\ref{eq:scaling_Pc}) and~(\ref{eq:scaling_Pqc}).
Thus the Gardner transition is closely related to the jamming scalings that appear due to {\it mechanical marginality}.
 In soft particles upon cooling, the {\it mechanical marginality} is absent, and the Gardner transition is expected due to {\it landscape marginality} of the free-energy landscape. The insight provided here would help to distinguish between the two types of marginality.  
 
 In recent studies, the jamming scalings Eq.~(\ref{eq:scaling_Pc}) and~(\ref{eq:scaling_Pqc}) have been verified by high-precision force and gap measurements in athermal photo-elastic granular disks near the jamming transition~\cite{wang2022experimental, shang2024yielding}. Along this line,  we propose to detect the structural signature and  order parameter of the Gardner transition in future granular experiments, by implementing the vibrational setup to mimic thermal hard disks~\cite{seguin2016experimental}.

{\bf Acknowledgment.}
We warmly thank  Patrick Charbonneau for inspiring discussions. 
We acknowledge financial support from NSFC (Grants 12474189, 1240042251, 12161141007, 11935002, 52394163, 12047503 and 12404290), from 
Chinese Academy of Sciences (Grant ZDBS-LY-7017), 
and from Wenzhou Institute (Grant WIUCASQD2023009). 
In this work access was granted to the High-Performance Computing Cluster of Institute of Theoretical Physics - the Chinese Academy of Sciences.

\clearpage 

{\bf \Large End Matter}
\vspace{0.5cm}

{\it Appendix A: Weighted bond orientational order parameters.}
The parameter $Q_6^{\rm w}$ in 3D~\cite{mickel2013shortcomings} is the mean value of local order parameters, $ Q_6^{\rm w} = \frac{1}{N} \sum_{i=1}^{N} Q_{6\rm,i}^{\rm w}$, where 
\beq
Q_{6\rm,i}^{\rm w} = \sqrt{\frac{4\pi}{13} \sum_{m = -6}^{6}\left|\sum_{j \in \partial i} \frac{A_{j}}{A} Y_{6m}(\theta_{j}, \phi_j)\right|^2},
\eeq
and $Y_{6m}(\theta,\phi)$ is the spherical harmonic function.
The $\Psi_6^{\rm w}$ in 2D is defined in a similar way~\cite{li2020attraction}, $\Psi_6^{\rm w} = \frac{1}{N} \sum_{i=1}^{N} \Psi_{6\rm,i}^{\rm w}$, where 
\beq
\Psi_{6\rm,i}^{\rm w} = \sqrt{
\left|\sum_{j \in \partial i} \frac{A_j}{A} {\exp(\mathbf{i} 6 \theta_j)}\right|^2}.
\eeq
The neighboring particles $i$ and $j$ are determined via the radical Voronoi tessellation method  implemented by the voro++ program~\cite{Rycroft2009voro++}.
In 3D, $A_{j}$ is the area of the $j$-th facet, $A$ is the total area of Voronoi polyhedron, $\theta_j$ and $\phi_j$ are the spherical angles. In 2D, $A_{j}$ is the length of the  $j$-th edge, $A$ is the perimeter of Voronoi polygon, and $\theta_j$ is the polar angle.\\

{\it Appendix B: Definition of inter-particle gaps.}
The inter-particle gaps between Voronoi neighboring particles, denoted by $x$, is calculated as the distance between two particles minus the sum of particle radii, $x = r_{ij} - (D_i+D_j)/2$.
The length unit is $\frac{L}{N^{1/d}}$ in $d$ dimensions, where $L$ is the linear size of the simulation box and $N$ is the total number of particles. Periodic boundary conditions are employed to compute the distance.
\\

\begin{figure}[!htbp]
  \centering
  \includegraphics[width=0.9\linewidth]{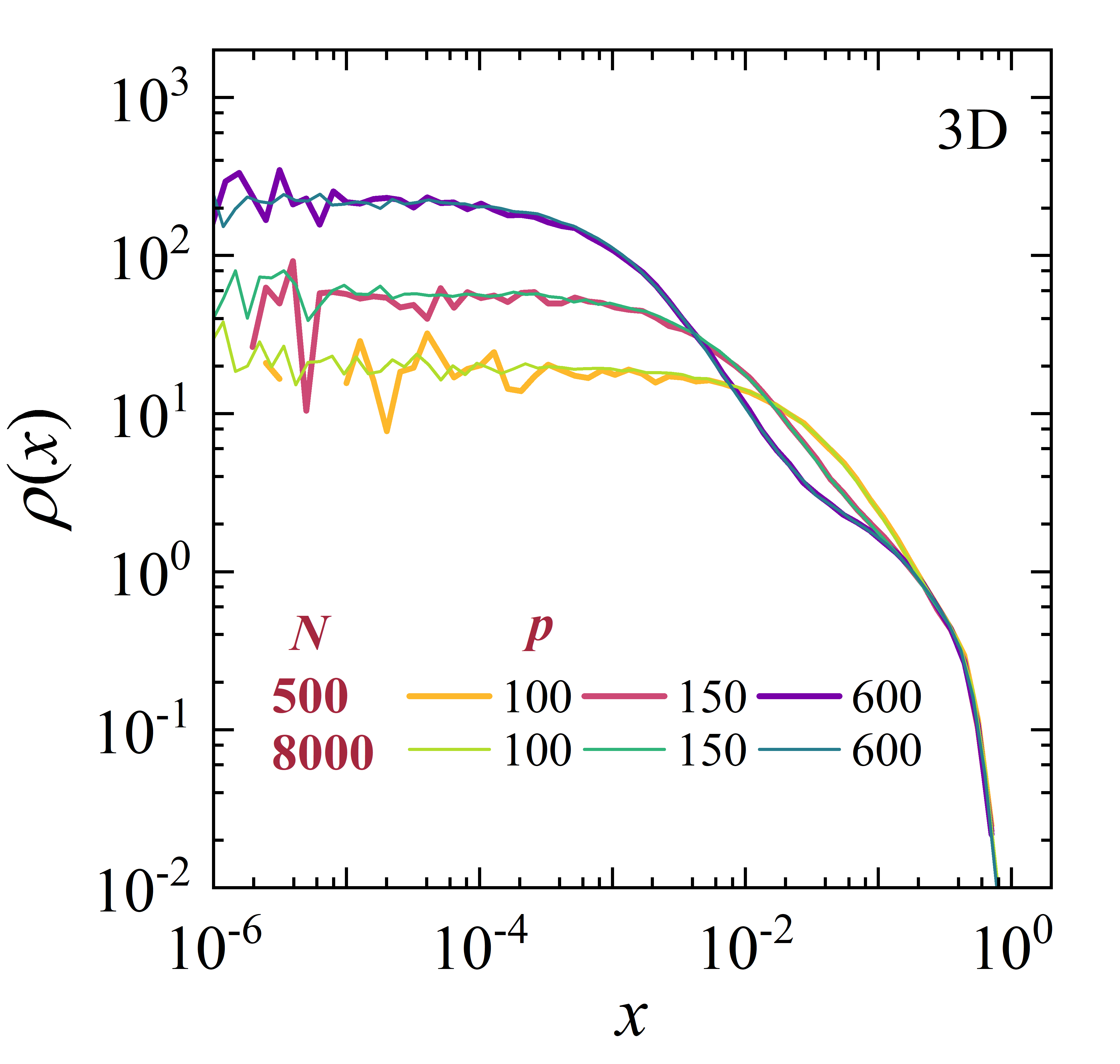}
  \caption{Gap distribution $\rho(x)$ for two different system sizes $N$, at three different $p$, in 3D.
  } 
\label{fig:differentN}
\end{figure}

\begin{figure}[!htbp]
  \centering
  \includegraphics[width=0.9\linewidth]{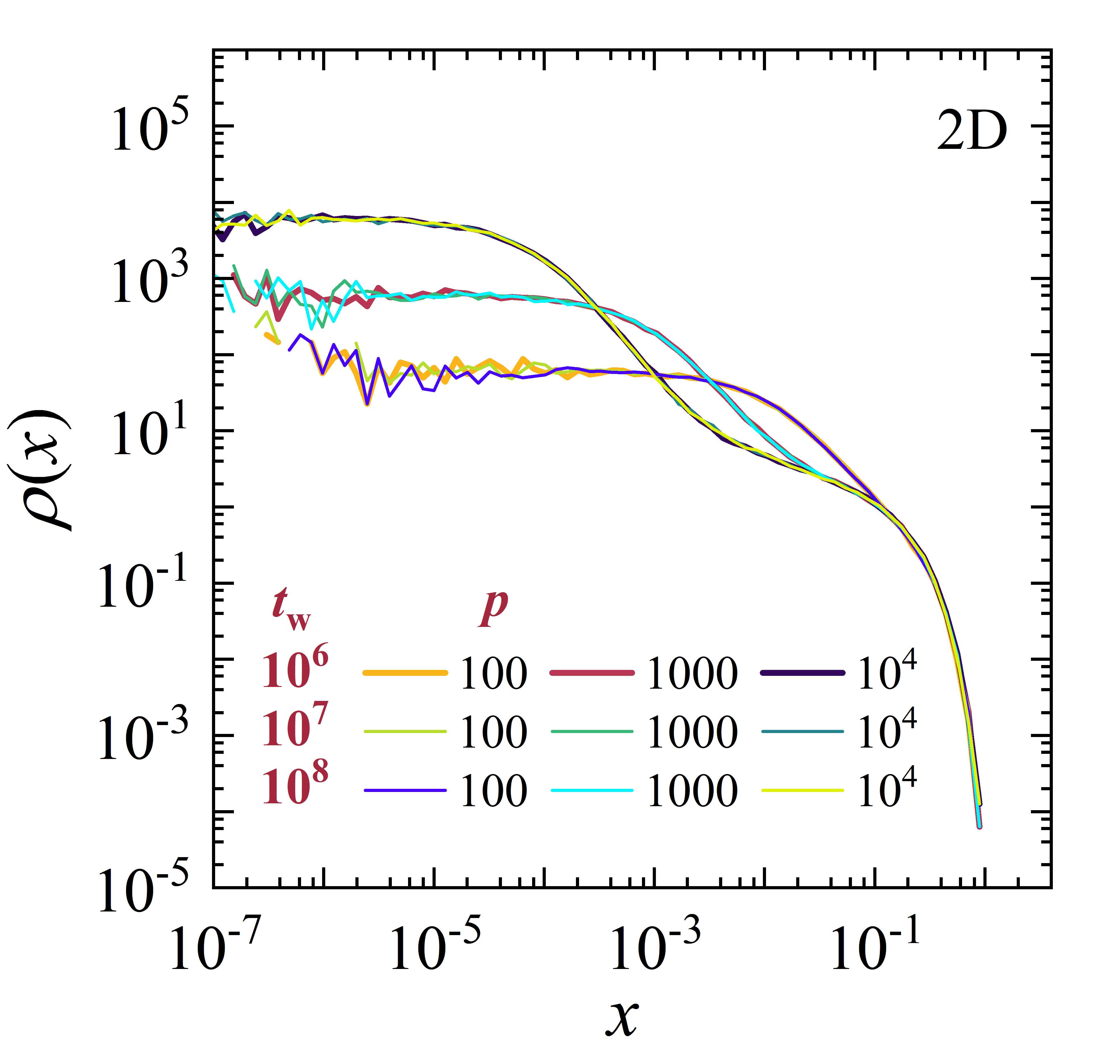}
  \caption{Gap distribution $\rho(x)$ for three different waiting times $t_{\rm w}$, at three different $p$, in 2D.
  } 
\label{fig:differenttw}
\end{figure}

{\it Appendix C: Models.}
The systems, in both 2D and 3D, consist of $N$ hard-core particles with a diameter  distribution  $P(D) \sim D^{-3}$ for $D_{\rm min} < D < D_{\rm min}/0.45$~\cite{berthier2016equilibrium,berthier2016growing,liao2019hierarchical,li2021determining}. 
The results in the main text are obtained for   $N = 1024$ (2D) and $N = 500$ (3D) systems, and the $N$-dependence is examined in {Appendix F}.
The dimensionless pressure $p$ is defined by 
$p = PV/(Nk_{\rm B}T)$, where $P$, $V$, $T$ and $k_{\rm B}$  are the pressure, volume, temperature and Boltzmann constant (we set $T=1$ and $k_{\rm B}=1$). The pressure $P$ is computed from binary collisions~\cite{lubachevsky1990geometric}.
The  particle mass is $m=1$.
We generate 48 independent samples in both dimensions to achieve good statistics, with 25 samples used for static structure analysis and 48 samples for dynamic analysis.
\\

{\it Appendix D: Molecular dynamics simulation algorithms.}
In 3D, samples are obtained by the  Lubachevsky–Stillinger algorithm with a constant inflating rate $\Gamma = \frac{1}{2D}\frac{dD}{dt} = 10^{-3}$ as in Ref.~\cite{berthier2016growing}. The time unit is $\left( \frac{1}{k_{\rm B}m}\right)^{1/2} \overline{D}$, where $\overline{D}$ is the mean diameter.

In 2D, the constant pressure discrete-time Langevin molecular dynamics method (an NPT-ensemble version of the Lubachevsky–Stillinger algorithm)~\cite{Grnbech2014Constant} is used to instantaneously  bring the system to the target pressure $p$. 
In order to check the time-dependence, a waiting time $t_{\rm w}$  is introduced after the instantaneous quench. 
The $t_{\rm w}$ is expressed in number of collisions in order to make it easier to be compared with previous Monte-Carlo (MC) simulations~\cite{liao2019hierarchical} (at $p=100$, one collision time is approximately equivalent to 100 MC steps in~\cite{liao2019hierarchical}).
The results in the main text are obtained for $t_{\rm w} = 10^6$ collisions, and the $t_{\rm w}$-dependence is examined in {Appendix F}.\\

\begin{figure}[!htbp]
  \centering
  \includegraphics[width=1.0\linewidth]{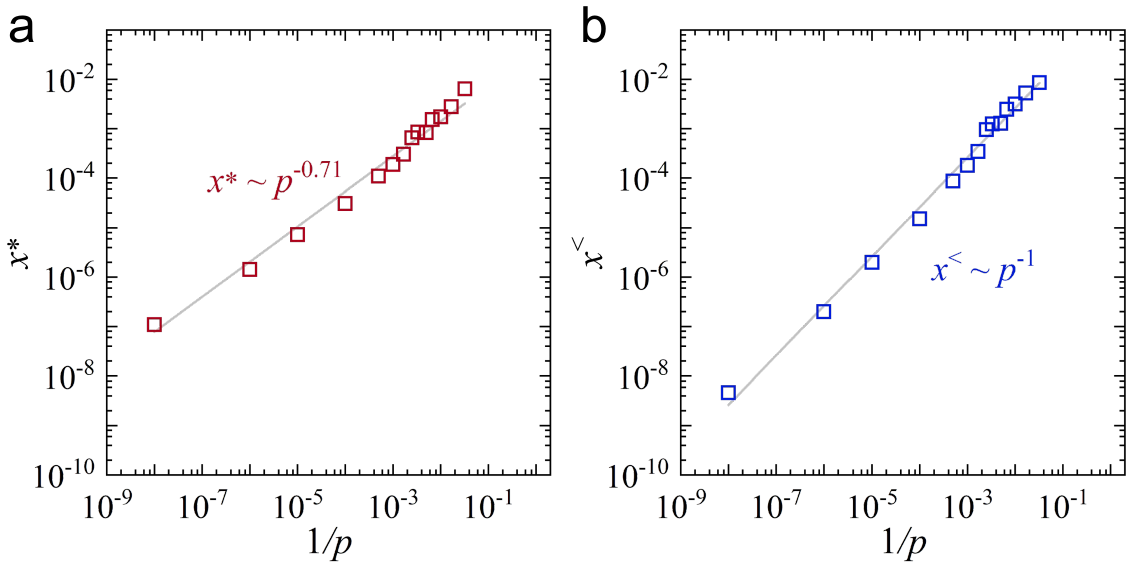}
  \caption{The numerically determined $x^*$ and $x^<$ as functions of pressure. The solid lines are the scaling relations (a) $x^* \sim p^{-0.71}$ and (b) $x^< \sim p^{-1}$.
  } 
\label{fig:x_critical_points}
\end{figure}

{\it Appendix E: Pressure-scaling of parameters $k$ and $a$ in Eq.~(\ref{eq:Peq}).} We fit the gap distribution $\rho(x)$ for the part of $x>x^*$ by the ``equilibrium'' form  Eq.~(1) with fixed $\beta = 0$ and  $\gamma = 0.8$ (see Fig.~\ref{fig:PDF}d). The remaining fitting parameters $k$ and $a$ are plotted in Fig.~\ref{fig:rhoeq_parameters} as functions of $p$. It shows that $k$ scales as $k \sim p$, and  $a$ sales as $a \sim p^{0.07}$.\\

{\it Appendix F: Independence of the gap distribution on  the system size $N$ and waiting time $t_{\rm w}$.} Figures~\ref{fig:differentN} and~\ref{fig:differenttw} suggest that $\rho(x)$ is independent of 
$N$ and $t_{\rm w}$. Consequently, the structural order parameter should be insensitive to $N$ and $t_{\rm w}$.\\

{\it Appendix G: Pressure-scaling of $x^*$ and $x^<$.} We numerically estimate the two separation points $x^*$ and $x^<$ (see Fig.~\ref{fig:x_critical_points}). The two predicted scaling relations $x^* \sim p^{-0.71}$ and $x^< \sim p^{-1}$ are also shown in the figure.\\

\end{document}